\documentclass[aps,preprint,floatfix]{revtex4-1}
\usepackage{amsmath}
\usepackage{latexsym}
\usepackage{float}
\usepackage{amssymb}
\usepackage{graphicx}
\usepackage{textcomp}
\usepackage{hyperref}
\usepackage{color}
\newcommand{\bea}{\begin{eqnarray}}
\newcommand{\eea}{\end{eqnarray}}
\newcommand{\vect}[1]{\mathbf{#1}}
\newcommand{\req}{\rho_{\rm eq}}
\newcommand{\mex}{\mu^{\rm ex}}
\newcommand{\di}{\displaystyle}
\newcommand{\kt}{k_{\rm B}T}

\begin{document}

\title{Hard sphere fluids at a soft repulsive wall: A comparative study using Monte Carlo and density functional methods}
\author{Debabrata Deb, Alexander Winkler, Mohammad Hossein Yamani, Martin Oettel, Peter Virnau, and Kurt Binder}
\affiliation{Institut f\"ur Physik, Johannes Gutenberg-Universit\"at Mainz, \\ Staudinger Weg 7, 55099 Mainz}
\begin{abstract}
Hard-sphere fluids confined between parallel plates a distance $D$ apart are studied for a wide range of packing fractions, including also the onset of crystallization, applying Monte Carlo simulation techniques and density functional theory. The walls repel the hard spheres (of diameter $\sigma$) with a Weeks-Chandler-Andersen (WCA) potential $V_{WCA}(z) = 4 \epsilon [(\sigma_w/z)^{12}-(\sigma_w/z)^6 + 1/4]$, with range $\sigma_w = \sigma/2$. We vary the strength $\epsilon$ over a wide range and the case of simple hard walls is also treated for comparison. By the variation of $\epsilon$ one can change both the surface excess packing fraction and the wall-fluid $(\gamma_{wf})$ and wall-crystal $(\gamma_{wc})$ surface free energies. Several different methods to extract $\gamma_{wf}$ and $\gamma_{wc}$ from Monte Carlo (MC) simulations are implemented, and their accuracy and efficiency is comparatively discussed. The density functional theory (DFT) using Fundamental Measure functionals is found to be quantitatively accurate over a wide range of packing fractions; small deviations between DFT and MC near the fluid to crystal transition need to be studied further. Our results on density profiles near soft walls could be useful to interpret corresponding experiments with suitable colloidal dispersions.
\end{abstract}
\maketitle

\section{Introduction}
Recently it has been demonstrated that for colloidal suspensions the effective interactions are tunable from hard spheres to soft repulsion \cite{1,2,3} or weak attraction \cite{4,5,6}, and at the same time the structure of fluid-crystal \cite{7,8,9,10} and fluid-wall interfaces can be analyzed in arbitrary detail, e.g. by visualizing the packing of particles in these interfaces \cite{10}. Correspondingly, there is a great interest in model studies pertinent to such systems. However, most work has focused on the archetypical hard sphere fluid \cite{11,12,13,Dellago11}, confined by hard walls \cite{14,15,16,17,18,19,20,21,22,23,24,25,26,27,28}. With respect to heterogeneous crystal nucleation at hard walls \cite{22,23}, this system is difficult to understand, since there is evidence that complete wetting of the wall by the crystal occurs, when the fluid packing fraction approaches the fluid-crystal phase boundary in the bulk \cite{23}.

Now it is well known that the interaction between colloidal particles and walls can also be manipulated, by suitable coatings of the latter, e.g. via a grafted polymeric layer (using the grafting density and chain length of these polymers, under good solvent conditions, as control parameters \cite{29,30}). Thus, in the present work we explore a model where colloidal particles that have an effective hard-sphere interaction in the bulk experience a soft repulsion from confining walls, describing this repulsion for the sake of simplicity by the Weeks-Chandler-Andersen \cite{31} potential. We show that such a short-range repulsion has only small effects on the structure of the fluid near the wall, but nevertheless affects the wall-fluid interface tension $\gamma_{wf}$ significantly. Both Monte Carlo methods and density functional calculations are used.

In Sec.~2, the model is introduced, and several Monte Carlo methods to extract $\gamma_{wf}$ are briefly described. Since the judgment of accuracy for such methods is somewhat subtle \cite{27}, we are interested in comparing estimates from several rather different approaches, to avoid misleading conclusions. In Sec.~3, we present our results on density profiles, while Sec.~4 describes our results for the dependence of $\gamma_{wf}$ on packing fraction. First preliminary results on the interfacial tension between the crystalline phase and the confining wall are presented in Sec.~5, while Sec.~6 summarizes our results and discusses possible applications to experiments. The density functional methods are briefly explained in an Appendix.

\section{Model and summary of Monte Carlo methods for the estimation of wall free energies}
The simulated model is the simple fluid of hard particles of diameter $\sigma$, in the geometry of an $L \times L \times D$ system, confined between two parallel walls located at $z=0$ and at $z=D$. In the x and y directions, periodic boundary conditions are applied throughout. The particle-wall interaction contains either a hard wall type interaction

\begin{equation}\label{eq1}
V_{HW}(z)= \infty \quad \textrm{for} \quad z < \sigma /2 \quad \textrm{and for} \quad z >D-\sigma /2
\end{equation}

or a soft repulsion of the Weeks-Chandler-Andersen \cite{31} type


\begin{equation}\label{eq2}  
  \begin{array}{lll}
    V_{WCA}(z) & = 4 \epsilon \left[(\sigma _w/z)^{12} - (\sigma _w/z)^6 + \frac{1}{4}\right] & \mbox{for } 0\leq z \leq \sigma_w2^{1/6}\\ 
               & = 4 \epsilon \left[(\sigma_w/(D-z))^{12}-(\sigma _w/(D-z))^6+\frac 1 4\right] & \mbox{for } (D-\sigma_w2^{1/6}) \leq z  \leq D\\
               & = 0 & \mbox{otherwise}
  \end{array}
\end{equation}


In Eq.~\ref{eq2} we choose $\sigma_w= \sigma /2$, while the parameter $\epsilon$ that controls the strength of this additional soft repulsion is varied in the range $0 \leq \epsilon \leq 4$ (choosing units such that $k_B =1$ and absolute temperature $T=1$). Note that for $\epsilon =0$ Eq.~\ref{eq2} also becomes a hard-core potential $V_{HC}(z)=\infty$ for $z<0$ and $z>D$, respectively.

In the literature the confinement of hard spheres between hard walls. i.e. the case where only $V_{HW}(z)$ is present, has already been extensively studied \cite{14,15,16,17,18,19,20,21,22,23,24,25,26,27,28}, while we are not aware of any work using $V_{WCA}(z)$ instead. The advantage of the choice Eq.~\ref{eq2} from the theoretical point of view, is that $\epsilon$ is a convenient control parameter: varying $\epsilon$ the wall-fluid interfacial tension $\gamma_{wf}$ as well as the wall-crystal interfacial tension $\gamma_{wc}$ can be modified. Note that the direct effect of $V_{WCA}(z)$ is zero in the range $\sigma_w2^{1/6}<z<D-\sigma_w2^{1/6}$: thus, when $D$ is very large, we expect that the structure of the hard sphere fluid in the center of the slit (very far from both walls) is identical to a corresponding hard sphere fluid in the absence of confining walls (applying periodic boundary conditions also in the $z$-direction).

We stress that the WCA form of the potential in Eq.~\ref{eq2} is only chosen for the sake of computational convenience. Having the application to colloidal dispersions in mind, one might expect that the colloidal particles carry a weak electrical charge, but the Coulomb interactions are strongly screened by counterions in the solution. Assuming also some effective charges at the walls, a potential like $C \exp(-\kappa z)$ might seem a physically more natural choice (with a screening length $\kappa ^{-1}$ of the order of $\sigma/10$ \cite{3} or even smaller; note that the constant $C$ could be positive or negative). However, when flexible polymers are adsorbed (or end-grafted) at the walls, the chain length $N$ and grafting density $\sigma_g$ provide additional parameters of a repulsive potential due to the dangling chain ends out in the solution, if there is no adsorption of the polymers on the colloidal particles. Thus, the actual potential between colloidal particles and confining walls is clearly non-universal, it depends on system preparation and can be fairly complicated due to a superposition of several mechanisms. Since we do not attempt to model any specific system, we take Eqs.~\ref{eq1},\ref{eq2} as a generic model.

For simulations in the standard canonical (constant volume) ensemble, the standard Monte Carlo algorithm \cite{32} with local single particle moves is implemented, choosing particles at random and attempting to move their center of mass to a new position. Of course, moves are accepted only if they respect the excluded volume between the particles. For the system with walls, the Metropolis criterion needs to be tested if either the old or the new position of the particle is within the range of the wall potential, Eq.~\ref{eq2}. At this point, the advantage of choosing a potential that is strictly zero for a broad range of $z$ (as specified above) clearly becomes apparent.

The observables of interest (for simulations in the canonical ensemble) are the normal pressure $P_N$ and the local tangential pressure $P_T(z)$ and the corresponding number density profile $\rho(z)$, choosing the average particle density $\rho = \int \limits ^D _0 \rho (z) dz /D=N/(L^2D)$ or the corresponding packing fraction

\begin{equation}\label{eq3}
\eta = (\pi \sigma ^3/6 ) \rho
\end{equation}

as the input parameters that we vary in our simulation.

Note that due to wall effects on the hard sphere fluid we expect an approach to the bulk density $\rho_b(P_N,T)$ as $D \rightarrow \infty$ as follows \cite{33}

\begin{equation}\label{eq4}
\rho = \rho_b(P_N,T) + 2 \rho_s/D, \quad D \rightarrow \infty
\end{equation}

where the surface excess density $\rho _s $ (and associated surface excess packing fraction $\eta_s$) are formally defined for a semi-infinite system as

\begin{equation}\label{eq5}
\rho_s= \int \limits _0^\infty [\rho(z)-\rho_b] dz \quad , \quad \eta_s = \rho_s \pi/6 \;.
\end{equation}

In a film of finite thickness $D$, an analog of Eq.~\ref{eq5} can be used if $\rho(z)$ has settled down to $\rho_b$ already for values of $z$ that are clearly smaller than $D/2$: then the upper limit $\infty$ in Eq.~\ref{eq5} can be replaced by $D/2$ with negligible error. In this limit, the two walls can be considered as strictly non-interacting, and then the wall-fluid interfacial tension is also \cite{26,33} simply related to the difference between $P_N$ and the average tangential pressure, $P_T= \int \limits_0^Ddz P_T(z)/D$,

\begin{equation}\label{eq6}
\gamma_{wf}= (P_N-P_T)D/2.
\end{equation}

Note, however, that the situation is more subtle for a crystal confined between two walls, since the long range crystalline order in the crystal is not necessarily commensurate with the chosen distance $D$ and hence the long-range elastic distortion of the crystal that will in general result invalidates the above statement that the effects of the two walls add independently. But, for fluid systems Eq.~\ref{eq6} is useful if $D$ is large enough.

As is well known, the standard ``mechanical'' approach to calculate the pressure from the virial expression \cite{33,34,35} cannot be straightforwardly applied for systems with hard-core interactions. In order to apply Eq.~\ref{eq6}, we thus follow the approach of de Miguel and Jackson \cite{26}. We here recall only briefly the most salient features. For a bulk hard sphere fluid the number of pairs with a relative distance in the range from $\sigma$ to $\sigma + \Delta r$ is sampled, $n(\Delta r)$, and one estimates the derivative of this function for $\Delta r \rightarrow 0$, $a \equiv \sigma d (n (\Delta r))/dr$, and uses the formula \cite{26}

\begin{equation}\label{eq7}
P/(\rho k_BT)=1+ a /(3N)
\end{equation}

to obtain the (average) pressure of a bulk hard sphere fluid at given density $\rho=N/V$. Alternatively, one can consider virtual volume changes by a factor $\xi$ and compute the probability $P_{nov}(\xi)$ that there are no molecular pair overlaps when the volume is decreased from $V$ to $V'=V(1-\xi)$. For small $\xi$ one can show that $P_{nov} (\xi)=\exp (-b\xi)$, where $b >0$  is related to the pressure by a relation similar to Eq.~\ref{eq7} \cite{26}

\begin{equation}\label{eq8}
P/(\rho k_BT)=1+b/N,
\end{equation}

and one can numerically verify that both routes based on Eqs.~\ref{eq7}, \ref{eq8} work in practice, and agree within their statistical errors. The method of Eq.~\ref{eq8} now can be straightforwardly extended to sample $P_N$ and $P_T$ separately: one considers volume changes that are due to reducing the distance from $D$ to $D'=D(1-\xi)$ keeping the lateral distance $L$ constant to obtain $P_N$, while $L'=L(1-\xi)$ at fixed $D$ is used to obtain $P_T$ \cite{26}.

When we vary the strength $\epsilon$ of the WCA \{Eq.~\ref{eq2}\} wall potential for fixed total particle number $N$ and fixed linear dimensions $L$ and $D$, the change of $\rho_s$ caused by the variation of $\epsilon$ necessarily cause a change of $\rho_b$ (and hence $P_N$), since in the canonical ensemble the total density is strictly constant. However, this effect is clearly undesirable: we want to vary $\epsilon $ and $\rho_s$ but keep the bulk conditions unchanged! Hence it would be preferable to vary $\epsilon$ and keep $P_N$ constant, rather than keeping the volume constant. But we do wish to keep $D$ constant as well. At first sight, one might conclude that these constraints are impossible to realize, since $P_N$ and $D$ are a pair of thermodynamically conjugate variables. However, Varnik \cite{36} has devised an iterative method, where only the area $A=L^2$ rather than the whole volume $V=L^2D$ is allowed to fluctuate, as it would happen in an $NPT$ ensemble. Applying this method (for details, see \cite{36})  one can realize a $NP_NDT$ ensemble, and this ensemble is indeed useful to implement the variation of $\epsilon$. However, due to the larger computational effort of this method our simulations were done in the canonical NLDT ensemble.

In some cases of interest it suffices to compute differences $\Delta \gamma = \gamma_{wf}(\epsilon)-\gamma_{wf}(\epsilon_o)$ only, rather than the values $\gamma_{wf}(\epsilon),\; \gamma _{wf}(\epsilon_0)$ individually. E.g., for $\epsilon_0=0$ our model reduces to the case of hard walls only, Eq.~\ref{eq1}, which has been studied extensively in the literature \cite{14,15,16,17,18,19,20,21,22,23,24,25,26,27,28}, and rather precise values of $\gamma_{wf}(\infty)$ are already available \cite{23,24}. Such differences $\Delta \gamma$ can, at least in principle, be found from a thermodynamic integration method based on linear response theory. We note that the thermodynamic potential can be written as (for $N \rightarrow \infty, D \rightarrow \infty)$

\begin{eqnarray}\label{eq9}
G(P_N,N,D,T)=-k_BT \ln \int d\vec{X}\exp \bigg\{-\beta \mathcal{H}_b (\vec{X})- \beta P_NL^2D - \nonumber \\
- \beta \epsilon L^2\int \limits _0^D \rho(z,\vec{X}) V'_{WCA}(z)dz \bigg\}
\end{eqnarray}

Here prefactors of the partition function that are unimportant for the following argument are already omitted. $\beta =1/k_BT$ and $\vec{X}$ stands for a point in the configuration space of the system (i.e., $\vec{X}$ is just the set of coordinates of all the center of masses of the hard spheres). By $\mathcal{H}_b (\vec{X})$ we denote the interaction among the hard spheres (i.e., $\exp [-\beta \mathcal{H}_b(\vec{X})]=0$ if any pair of hard spheres overlaps). The interaction with the WCA-potential has been written out explicitly, denoting $V_{WCA}(z)=\epsilon V'_{WCA}(z)$, and defining $\rho(z,\vec{X})$ as the particle density in the infinitesimal interval $[z,z +dz]$ for the configuration $\vec{X}$.

We now consider the derivative of $G$ with respect to $\epsilon$, to find (see also \cite{37} for a related treatment of a binary Lennard-Jones mixture) that this derivative just can be interpreted as the sum of $L^2 (\frac{\partial \gamma _{wf}}{\partial \epsilon}) $ for the two walls (which are identical). Hence

\begin{equation}\label{eq10}
\left(\frac {\partial \gamma_{wf}}{\partial \epsilon}\right)_{P_NDT} = \int \limits _0^{D/2} \langle \rho(z,\vec{X})\rangle_\epsilon V'_{WCA} (z) dz,
\end{equation}

where the notation $\rho(z,\epsilon) = \langle \rho(z,\vec{X})\rangle _\epsilon$ is used to emphasize that the statistical average $\langle \cdots \rangle_\epsilon$ is carried out in an ensemble where a wall potential $V_{WCA}(z)= \epsilon V'_{WCA}(z)$ is nonzero only for $0<z<2^{1/6}\sigma _w=0.5 \cdot 2^{1/6}\approx 0.561$, for our choice $\sigma_w=\sigma /2 = 1/2$. From Eq.~\ref{eq10} we realize that any change of $\gamma_{wf}$ due to the variation of $\epsilon$ can only be due to the fact that the product $\rho(z,\epsilon) V'_{WCA}(z)$ changes when $\epsilon$ is varied. Now differences $\Delta \gamma $ can be computed from

\begin{equation}\label{eq11}
\Delta \gamma = \int \limits _{\epsilon_0}^\epsilon \left(\frac {\partial \gamma_{wf}}{\partial \epsilon}\right)_{P_NDT} d \epsilon = \int \limits _{\epsilon_0}^\epsilon d \epsilon ' \int \limits _0^{D/2} \rho(z,\epsilon ') V_{WCA}' (z) dz
\end{equation}

For use in actual computations the application of Eq.~\ref{eq11} is subtle since one needs to record $\rho(z,\epsilon ')$ in the range $0 <z<2^{1/6} \sigma _w$ with very high precision. At this point, we draw attention to another version of thermodynamic integration (termed ``Gibbs-Cahn integration'' \cite{28}) which can also be implemented if only hard walls are present (Eq.~\ref{eq1}): there one uses $\rho_s$ \{Eq.~\ref{eq5}\} to study the variation of $\gamma_{wf}$ with the bulk density $\rho_b(P_N,T)$ of the system (see Eq.~\ref{eq:gibbs_ads} below).

However, in the present work we rather use another variant  of thermodynamic integration, which is briefly characterized below. This method (which we refer to as ``ensemble switch method'') is a variant of the method used by Heni and L\"owen \cite{18}, where one gradually switches between a system without walls, applying periodic boundary conditions throughout, described by Hamiltonian $\mathcal{H}_1(\vec{X})$, and a system (with the same particle number and volume) with walls, $\mathcal{H}_2(\vec{X})$, writing the total Hamiltonian as

\begin{equation}\label{eq12}
\mathcal{H}(\vec{X}) = (1-\kappa)\mathcal{H}_1(\vec{X})+ \kappa\mathcal{H}_2(\vec{X})
\end{equation}

where $\kappa \in [0,1]$ is the parameter that is varied for calculating the free energy difference between the systems (1,2). In a simulation $\kappa$ is typically discretized and the system is allowed to move from $\kappa_i$ to $\kappa_{i+1}$ or $\kappa_{i-1}$ with a Metropolis step. The free energy between the two states is given by $k_BT(\ln P(i) - \ln P(i-1, \; i+1))$ where $P(i)$ is the relative probability of residing in state $i$. As this probability varies considerably with $\kappa$, a variant of Wang-Landau sampling \cite{38,39} is employed to eventually simulate each state with equal probability. In this way we can sample free energy differences relative to the free energy of the system with periodic boundary conditions as a function of $\kappa$ (for technical details see \cite{40}). For $\kappa = 1$ the wall free energy then follows from
\begin{equation}\label{eq13}
\gamma_{wf}(\rho,T)= \lim _ {D \rightarrow \infty} \frac{\Delta F(D)}{k_BT2A}
\end{equation}
with $A$ being the area of the wall.

Note that also in this method for finite $D$ the density $\rho$ in the system with walls $(\kappa =1)$ differs from the corresponding system with periodic boundary conditions $(\kappa =0$) due to the surface excess density. Thus an extrapolation to $D\rightarrow \infty $ is necessary.

\section{Density profiles of hard sphere fluids confined between WCA walls}

\begin{figure}
\includegraphics[scale=0.5]{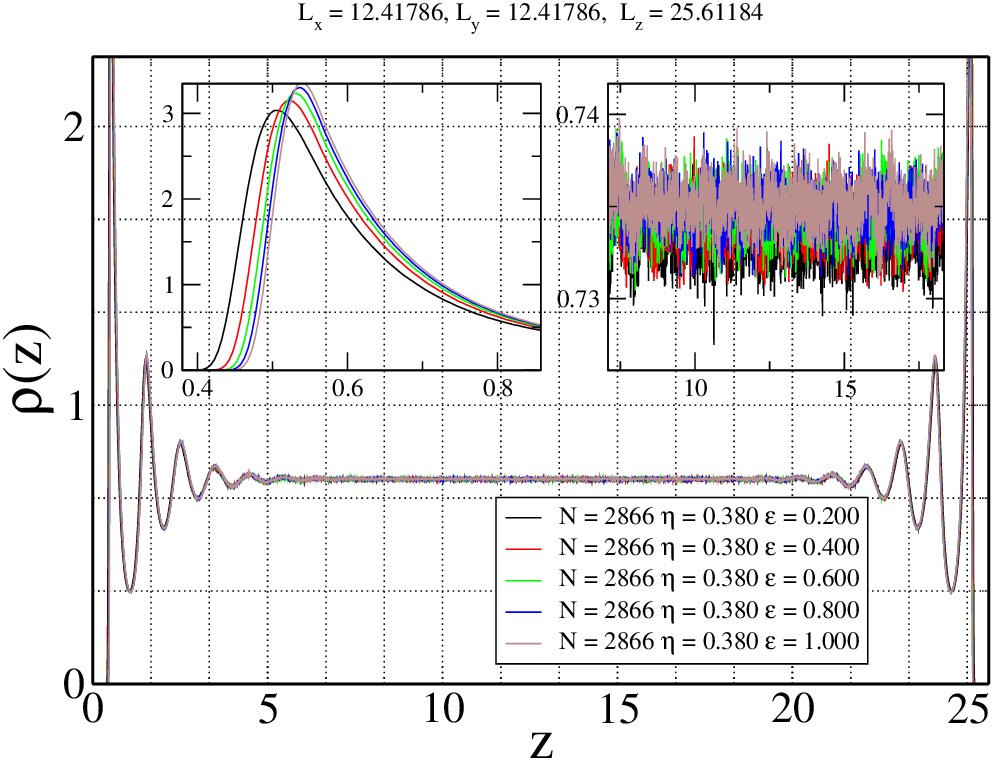}
\caption{\label{fig1} Density profile $\rho(z) $ vs. $z$, for a box of linear dimensions $L=12.41786, \; D=25.61184$, total particle number $N=2866$, and five choices of $\epsilon$, as indicated. The upper left inset shows the first peak of $\rho(z)$ close to the left wall, resolved on a much finer abscissa scale; the upper right inset shows the density in the central part of the box, resolved on a much larger ordinate scale, to show that for the different values of $\epsilon$ essentially the same bulk density $\rho_b$ in the center of the film is obtained.}
\end{figure}

\begin{figure}
\includegraphics[scale=0.5]{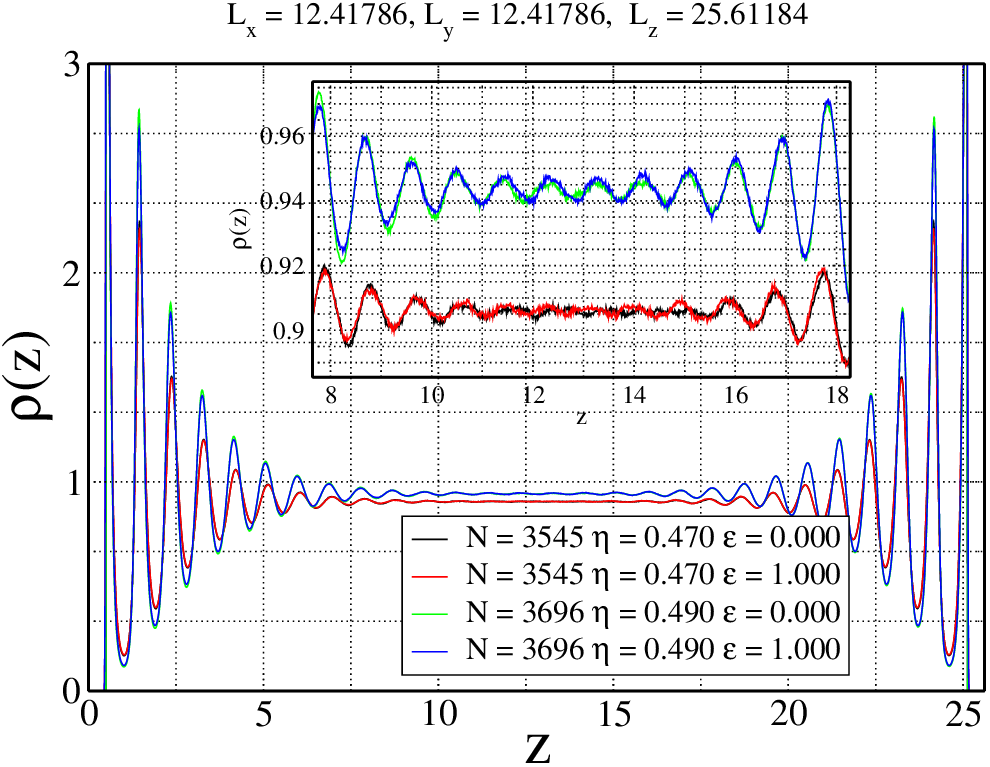}
\caption{\label{fig2} Same as Fig.~1, but for $N=3545$ and $N=3696$, respectively. Note that in both cases two choices of $\epsilon$ are shown, namely $\epsilon = 0$ (hard wall system) and $\epsilon = 1.0$, but on the scales of the plot these data coincide. Insert shows $\rho(z)$ in the center of the film on magnified scales, to show that at the densities chosen in this figure the distance $D$ chosen here is not large enough to render the two walls strictly noninteracting (the systematic density oscillations do not completely die out in the center of the film).}
\end{figure}

Figs.~\ref{fig1}, \ref{fig2} show typical data for the density profiles $\rho(z)$ obtained from our simulations, using a box of linear dimensions $L=12.41786, D=25.61184$, and varying the particle number $N$ as well as the strength  $\epsilon$ of the WCA potential, Eq.~\ref{eq2}.

At first sight, the density profiles for the different choices of $\epsilon$ look essentially identical; only when a magnified picture of the first peak of $\rho(z)$ adjacent to one of the walls is taken, one sees a systematic effect: the larger $\epsilon$, the more remote from the wall the peak occurs, as expected. However, for $\sigma_w2^{1/6}<z<D-\sigma_w2^{1/6}$, i.e. outside the range where the wall potential acts, the effect of varying $\epsilon$ is negligible. However, for packing fractions $\eta$ close to the value $\eta_{b,cr}$ where in the bulk crystallization starts to set in, $\eta_{b,cr}=0.492$, such as $\eta_b \approx 0.47$ or larger, the wall-induced oscillations in the density profile (``layering'') extend throughout the film (Fig.~\ref{fig2}). This observation indicates that the chosen thickness $D$, as quoted above, is not large enough to allow an approach very close to the transition, when one tries to disentangle the effects of the walls (as measured by $\rho_s$ or $\eta_s$, Eq.~\ref{eq5}, respectively) and $\rho_b$ \{Eq.~\ref{eq4}\} or $\eta_b$.

We have compared the values for the normal pressure $P_N$ and corresponding value of $\eta_b$ as function of the nominal packing fraction $(\eta)$ chosen in our simulations, for a range of values for $\epsilon$, the strength of the WCA potential at the walls to literature data \cite{26,27,28}. This shows that in the chosen range of $\eta_b$ the linear dimensions L and D chosen here are large enough to allow a meaningful estimation of $\eta_b$. Due to the surface excess of the density, there is a systematic discrepancy between $\eta_b$ (the packing fraction in the center of the thin film) and $\eta$ (the total packing fraction in the film).

\begin{figure}
\includegraphics[scale=0.5]{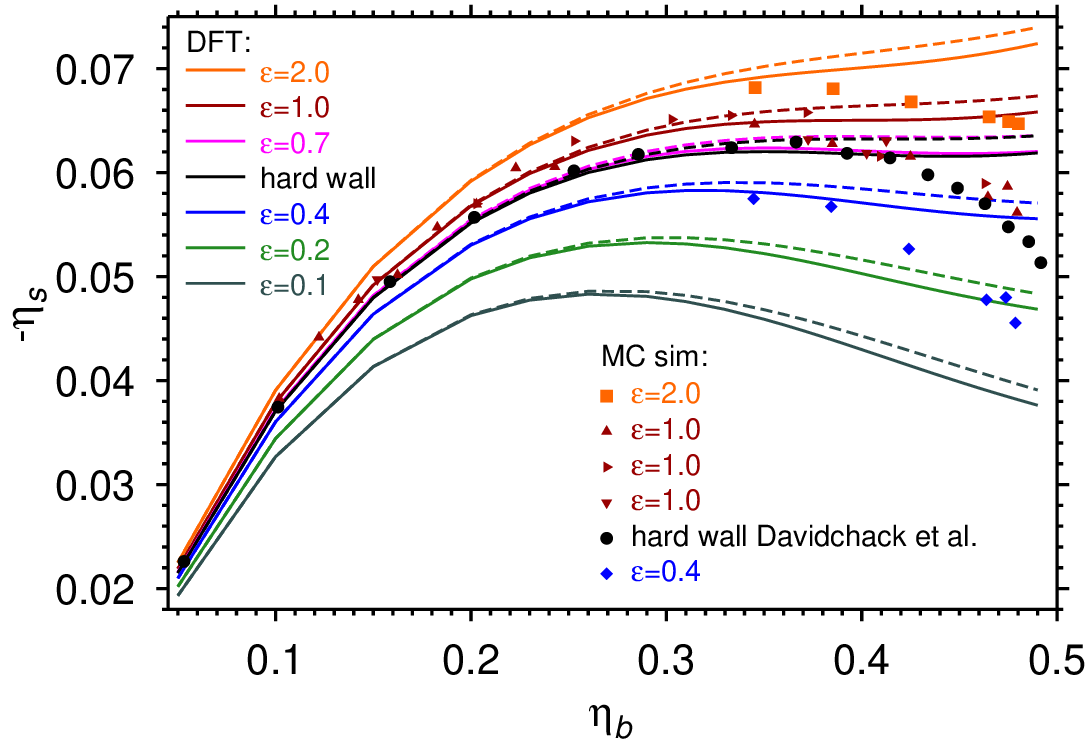}
\caption{\label{fig3} Plot of the surface excess packing fraction $(- \eta_s)$ versus the packing fraction $\eta_b=\rho_b \pi/6$, in the bulk for several choices of the strength $\epsilon$ of the WCA potential due to the wall (Eq.~2).
Some data are obtained from the same geometry as in Figs.~\ref{fig1}, \ref{fig2}, performing runs in the NVT ensemble (total density $\rho$ and corresponding packing fraction $\eta$ being held constant).
Data using other choices of $L$ and $D$ are included, to check for finite size effects. Triangles pointing to the right correspond to a geometry of $L=5$ and $D=40$. Triangles pointing down correspond to $L=13$ and $D=50$. All the other symbols correspond to a geometry of $L=12.418$ and $D=25.612$.
For comparison, also data for hard wall boundaries (Eq.~1) are included, both from the present work and from the data of Laird and Davidchack \cite{27,28} for the excess volume $v_N$ which can be related to the surface excess packing fraction as $- \eta_s = \rho_b v_N\pi/6$.
Symbols are Monte Carlo data, and lines show the the corresponding DFT results. Here, full curves
correspond to the White Bear II functional and broken curves correspond to the White Bear II (Tensor) functional.
The functionals differ in their applicability to describe the fluid--crystal transition (see appendix).}
\end{figure}

Fig.~\ref{fig3} shows a plot of $\eta_s$ (which turns out to be negative for all parameters that were studied) versus $\eta_b$. Corresponding results from the DFT calculations (see the Appendix for technical details) are included. One sees that $\eta_s$ depends in a nontrivial way on both $\eta_b$ and $\epsilon$. It can also be seen that for $\eta_{\textrm{bulk}}\approx 0.4$ systematic discrepancies between DFT and simulation start to occur, while for smaller $\eta_{\textrm{bulk}}$ both methods are in excellent agreement. Interestingly, for $\epsilon = 1$ the data are rather close to the case where a hardcore potential is used at the walls (data labeled as HW in Fig.~\ref{fig3}). The latter case has been studied before by Laird and Davidchack \cite{28}, and the present calculation is found to be in excellent agreement with these recent results. This very good agreement is rather gratifying, since the latter authors have studied a much larger system $(D = 65 \sigma, \; L = 50 \sigma$) than we have used. However, such larger systems are needed very close to the liquid-solid transition, due to the extended range of the layering (Fig.~\ref{fig2}). It is also suggestive that the behavior of $\eta_s$ for $\epsilon \rightarrow 0$ is singular (this limit again corresponds to the hard wall case, but a hard wall at a position shifted by $\sigma/2$). Note that the choice of the square cross section of the box (together with the periodic boundary condition) does not lead to noticeable systematic errors. Computations with a rectangular $L_x \times L_y$ cross section (compatible with a perfect triangular lattice of close-packed planes parallel to the walls) have also been made, but the results agree with those that are shown within the size of the symbols.

\begin{figure}
\includegraphics[scale=0.5]{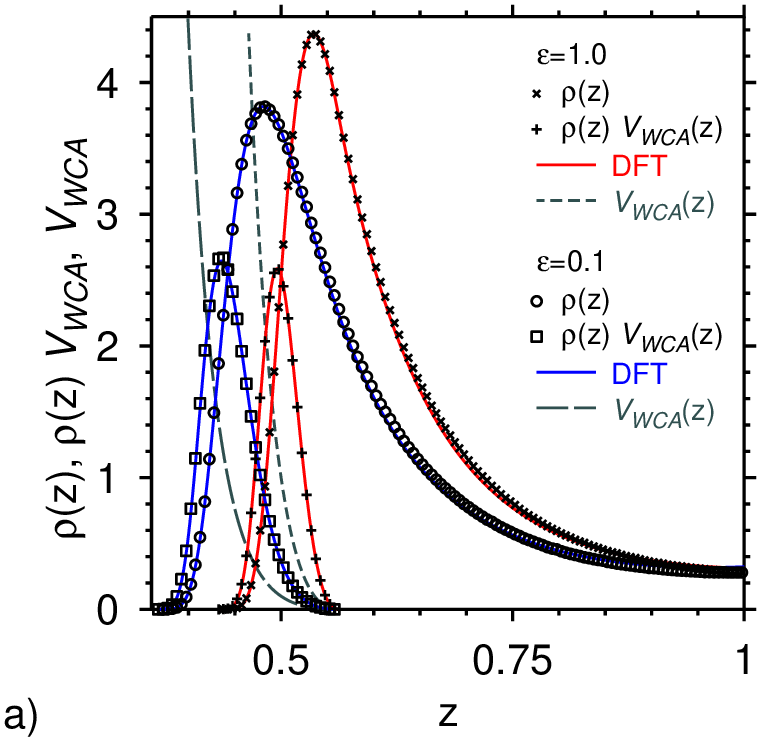}
\hspace{10 mm}.
\includegraphics[scale=0.5]{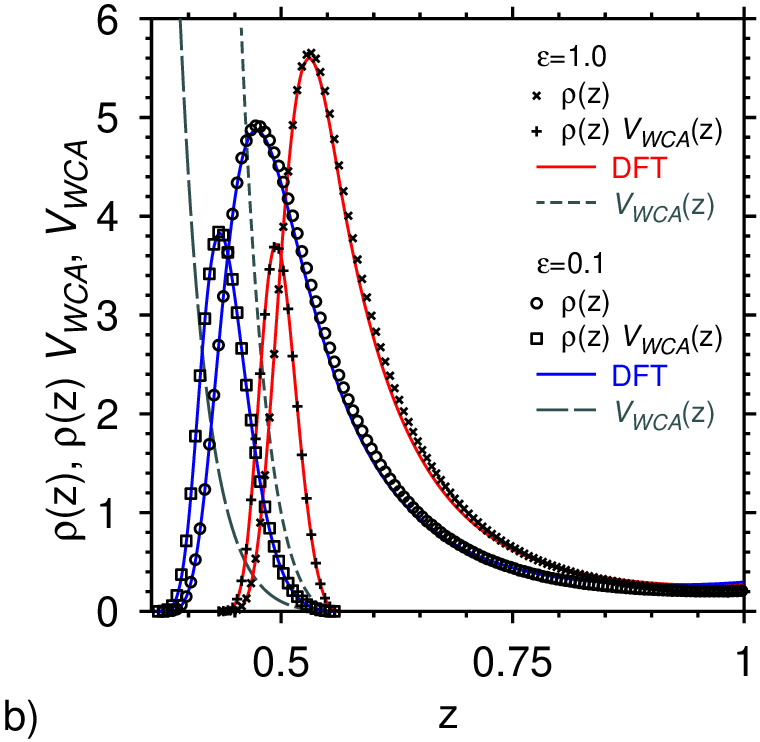}
\caption{\label{fig4} Wall potential $V_{WCA}(z)$, density $\rho(z)$ and product $V_{WCA}(z)\rho(z)$ plotted vs. $z$, in the regime $0.36 \leq z \leq 0.63$, for both $\epsilon=0.1$ and $\epsilon = 1.0$, for the case $\eta_b =0.42476$ (a) and $\eta_b = 0.46443$ (b). Density functional results (using the White Bear II functional)} are included, as full curves.
\end{figure}

\begin{figure}
\includegraphics[scale=0.5]{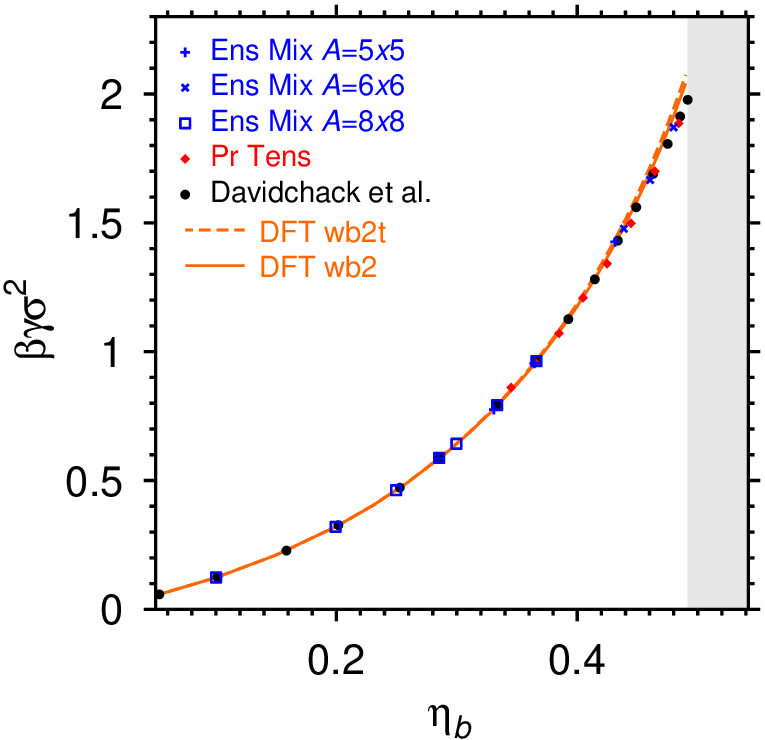}
\caption{\label{fig5} Wall-fluid surface tension $\gamma_{wf}$ of the hard sphere fluid confined by hard walls plotted vs. packing fraction $\eta_b$ in the bulk. Symbols show literature data \cite{26,27,28} and present results, due to the use of Eq.~\ref{eq6} and the thermodynamic integration method based on Eq.~\ref{eq12}, respectively; lines show the result of our DFT calculation (full lines -- White Bear II functional, broken lines --  White Bear II (Tensor) functional). Note that a factor $\sigma^2/k_BT$ is put equal to unity in this figure and the following figures throughout.}
\end{figure}

The distinct effect of the variation of $\epsilon$ on $\eta_s$ seen in Fig.~\ref{fig3} can already be taken as an indication that a clear effect on the interfacial tension $\gamma_{wf}$ can also be expected. To elucidate this point further, we present in Figs.~\ref{fig4}, \ref{fig5} in more detail the behavior of both $\rho(z)$ and $\epsilon \rho (z,\epsilon)V'_{WCA}(z)$. Recall that the product $\rho(z,\epsilon ')V'_{WCA}(z)$ appears in the integral when we relate $\gamma(\eta_b,\epsilon_0)$ and $\gamma(\eta_b;\epsilon)$ by thermodynamic integration \{Eq.~\ref{eq11}\}. Indeed one can see that the functions $\rho(z,\epsilon)V_{WCA}(z)$ does depend on $\epsilon$ significantly.

However, it is also clear from Fig.~\ref{fig4} that the use of Eq.~\ref{eq11} for practical computations would be difficult, since a very fine resolution of the z-dependence is necessary (while $z$ varies in between $0 < z < D=25.61184$, the important intervals contributing to Eq.~\ref{eq11} have only a width $\Delta z \approx 0.1$, and the location where these important intervals occur depend on $\epsilon$ and are not known precisely beforehand). Nevertheless the data of Figs.~\ref{fig3}, \ref{fig4} show that varying $\epsilon$ does have a pronounced effect on both the surface excess density $\rho_s$ (or packing fraction $\eta_s$, respectively) and on the function $V_{WCA}(z)\rho(z)$, and hence it is clear that varying $\epsilon$ must lead to a change of $\gamma_{wf}$ as well. This will be explored in the next section. It is also very gratifying that with respect to $V_{WCA}(z)\rho(z)$, the quantity that controls the surface tension $\gamma_{wf}(\epsilon)$, there is excellent agreement between the MC estimates and the DFT calculations for a wide range of packing fractions $(0 <\eta_b\leq 0.45$). Only in the immediate vicinity of the freezing transition $(0.46 \leq \eta_b \leq \eta_f$, with \cite{13,41} $\eta_f=0.492$) slight but systematic deviations are apparent in Fig.~\ref{fig4}b. For the surface excess density $\rho_s$, however, which is sensitive to the whole profile $\rho(z)$ and not only to the peaks of $\rho(z)$ next to the walls, deviations between DFT and MC start at smaller $\eta_b$ already. We add the caveat, however, that close to freezing the finite size effects on the density profile $\rho(z)$ need to be carefully studied (see the discussion of Fig.~\ref{fig2}) but this is left to future work.

\section{Surface free energies of the hard sphere model in the fluid phase}
As a test of our MC procedures, it is useful again to consider the hard wall case \{Eq.~\ref{eq10}\} first, since this case has been extensively studied in the literature \cite{18,26,27,28}. Fig.~\ref{fig5} gives evidence that our methods (based on Eq.~\ref{eq6} or Eq.~\ref{eq12}, respectively) are in mutual agreement and in agreement with the calculations in the literature, within the statistical errors expected for these data. Again the DFT calculation is in excellent agreement over a wide range of packing fractions with the simulation results. Only close to the freezing transition small but systematic deviations are present,
as can be expected from the differences in the surface excess density $\rho_s$ close to freezing (see Fig.~\ref{fig3}).
The surface excess density  and the surface tension $\gamma_{wf}$
  are connected through the Gibbs adsorption relation
\begin{equation}
 \label{eq:gibbs_ads}
 \rho_s = - \frac{\partial \gamma_{wf}}{\partial \mu_b} = - \rho_b \frac{\partial \gamma_{wf}}{\partial P_N(\rho_b)}\;,
\end{equation}
where $\mu_b$ and $P_N$ are the chemical potential and the bulk (normal) pressure, respectively,  pertaining to
the bulk density $\rho_b$.

\begin{figure}
\includegraphics[scale=0.5]{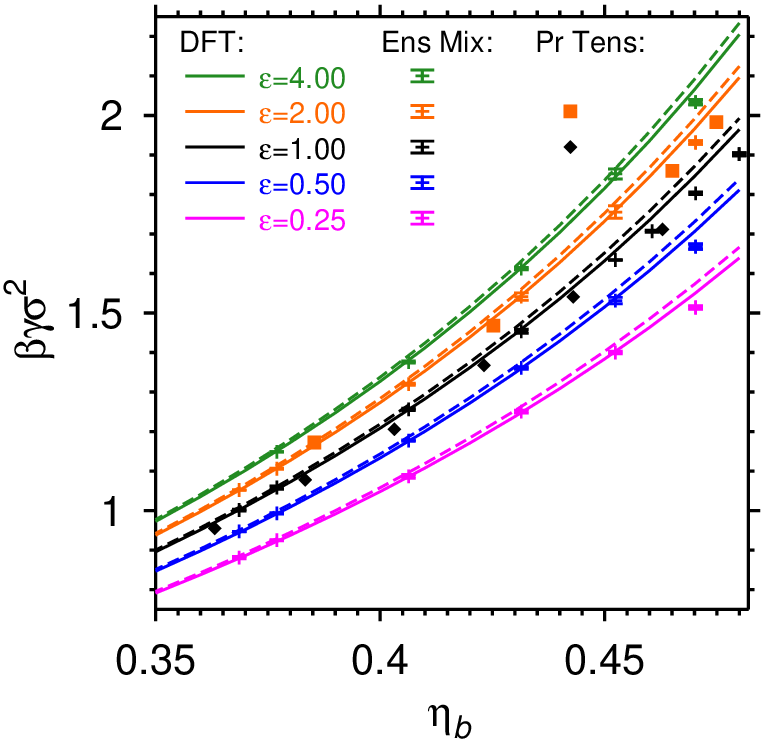}
\caption{\label{fig6} Wall-fluid surface tension $\gamma_{wf}$ plotted vs. packing fraction $\eta_b$, for the WCA wall potential \{Eq.~\ref{eq2}\}, varying its strength $\epsilon$ from $\epsilon=0.25$ to $\epsilon=4$, as indicated. Symbols are MC data,
 lines show the result of our DFT calculation (full lines -- White Bear II functional, broken lines --  White Bear II (Tensor) functional).}
\end{figure}

Having asserted that the errors of our calculations are reasonably under control, for the standard hard wall case, we turn to the problem of main interest in the present work, namely the variation of $\gamma_{wf}(\epsilon)$ with the strength $\epsilon$ of the WCA potential (Fig.\ref{fig6}). As we had expected, by changing $\epsilon$ we can indeed obtain a variation of $\gamma_{wf}(\epsilon)$ over a wide range. It is slightly disturbing, however, that there seem to be slight but systematic discrepancies between the MC results obtained from Eq.~\ref{eq6} and those from the thermodynamic integration method, Eq.~\ref{eq12}; this shows that the judgment of systematic and statistical errors in these methods is somewhat subtle. However, if we allow for statistical errors of the order of three standard deviation rather than one standard deviation, there would no longer be any significant discrepancy. Since for most purposes such a moderate accuracy in the estimation of $\gamma_{wf}(\epsilon)$ is good enough, we have not attempted to significantly improve the accuracy of our simulations, since this would require a massive investment of computer resources. Finally, we note that again the DFT results are very close to the MC data, particularly for $\eta_b \leq 0.45$ while closer to the freezing transition small but systematic discrepancies occur again. This very good agreement between DFT and simulations for $\gamma_{wf}$ is expected from the fact that DFT describes the density very accurately close to the walls where $V_{WCA}$ acts. More prominent deviations in the density profiles between simulation and DFT are seen near the second peak from the wall. DFT does not seem to account for its precise shape near freezing. This deficiency is also visible in the ``hump'' in the second peak of the pair correlation function near freezing which can be interpreted as a structural precursor to the freezing transition \cite{Tru98}.

\section{Some results on the wall-crystal surface tension}
As has already been stated earlier, studying the wall-crystal surface free energy is a subtle matter, since (i) in general there is always a misfit in a thin film geometry between the distance $D$ between the walls, and the lattice spacing $a(\eta_b)$ which depends on the packing fraction in the bulk, of course. In addition (ii) the wall-crystal free energy depends on the orientation of the crystal axes relative to the walls. In the present context, it is natural to restrict attention to a crystal orientation only where the close packed (111) planes at the face-centered cubic crystal structure (remember that in the fcc-structure there is an ABCABC... stacking of close-packed planes having a perfect triangular crystal structure each) are parallel to the planes forming the walls. Of course, it is this crystal orientation which occurs in wetting layers at the freezing transition from the fluid phase at the walls (if complete wetting at the transition occurs).

Thus, we have chosen values of $D$ such that the thickness is compatible with an integer number of stacked (111) lattice planes without creating a noticeable elastic distortion of the crystal. Only the thermodynamic integration method based on Eq.~\ref{eq12} is used, and system linear dimensions $L_x \times L_y \times D$ are taken, with $L_x \times L_y = [(8.8723 \times 7.6835),(8.8331 \times 7.6496),(8.7807 \times 7.6043),(8.7290 \times 7.5595),(8.6779 \times 7.5152),(8.6292 \times 7.4730)]$ and several choices of $D$, corresponding to $n = 6, 12, 24, 48$ stacked lattice planes. The result for $\gamma_{wc}$ does depend on $D$ but is compatible with a linear variation in $1/D$. So we find $\gamma_{wc}$ from an extrapolation versus $1/D\rightarrow 0$. We have checked the reliability of this approach for the case of the hard wall potential, Eq.~\ref{eq1}, where previous work with different methods have given \cite{24} $\gamma_{wc}^{111}(\eta_t)=1.457 \pm 0.018$.

\begin{figure}
\includegraphics[scale=0.5]{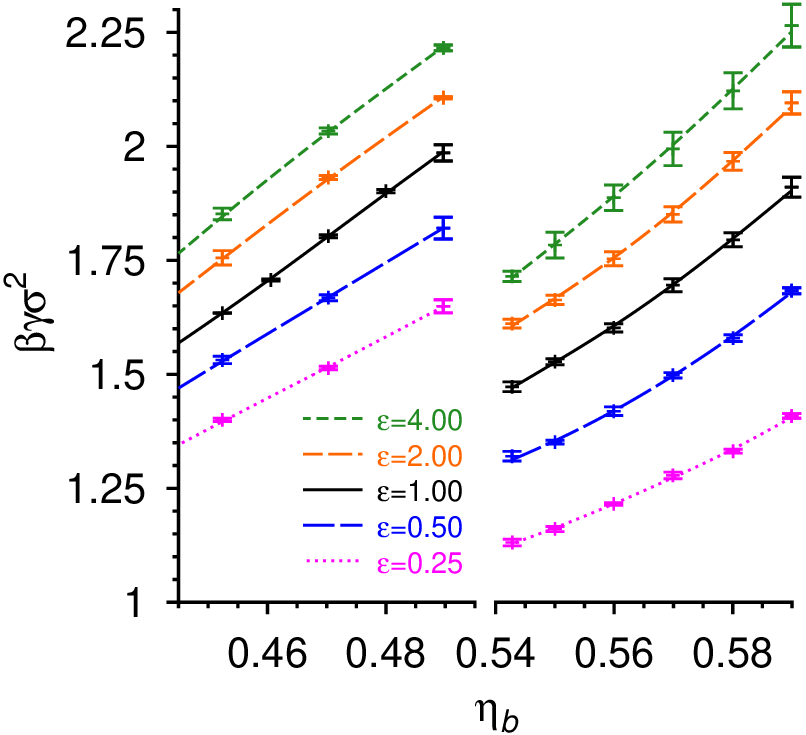}
\caption{\label{fig7} Wall-fluid surface tension $\gamma_{wf}$ and wall-crystal surface tension $\gamma_{wc}$ plotted vs. packing fraction, for the WCA potential \{Eq.~\ref{eq2}\}, for conditions near the fluid-solid transition. As in Fig.~\ref{fig6}, the strength $\epsilon$ of the WCA potential is varied: $\epsilon = 0.25, 0.5, 1.0, 2.0, $ and 4.0 (from bottom to top). The symbols are MC data obtained from the thermodynamic integration method based on Eq.~\ref{eq12}, error bars are from the linear fit as function of $D^{-1}$. At $\eta = 0.4896$ the smallest system size was excluded from the linear extrapolation since a crystalline layer was visible at the wall when looking at snapshots.}
\end{figure}

Fig.~\ref{fig7} shows our results for $\gamma_{wc}^{111}(\eta)$ for the WCA potential as a function of packing fraction and several choices of $\epsilon$. The corresponding data for $\gamma_{wf}$ for the fluid near the transition are also included. We find that the choice $\epsilon = 1$ yields functions $\gamma_{wf}(\eta), \; \gamma_{wc}(\eta)$ which are very close to the corresponding data for the hard wall case. For the latter, Fortini and Dijkstra \cite{24} have found that $\gamma_{wf}(\eta_t) = 1.990 \pm 0.007$, and hence the difference $\gamma_{wf}(\eta_t)-\gamma_{wc}^{111}=0.53 \pm 0.02$. Laird and Davidchack \cite{27} find $\gamma_{wf}(\eta_t) = 1.975 \pm 0.002$ and the difference $\gamma_{wf}(\eta_t)-\gamma_{wc}^{111}=0.563 \pm 0.004$. These difference values are very close to the fluid-crystal interface tension. Recent simulations by Laird and Davidchack using the cleaving method give \cite{27}  $\gamma^{111}=0.557 \pm 0.007$, $\gamma^{110}=0.571 \pm 0.006$ and $\gamma^{100} = 0.592 \pm 0.007$. The capillary wave fluctuation method, applied by the same authors, gives \cite{27a} $\gamma^{111}=0.546 \pm 0.016$, $\gamma^{110}=0.557 \pm 0.017$ and $\gamma^{100} = 0.574 \pm 0.017$ and the most recent simulation results from 2010 \cite{davidchack:2010} yield $\gamma^{111} = 0.5416$, $\gamma^{110} = 0.5590$ and $\gamma^{100} = 0.5820$ with uncertainties in the last two digits. These results imply that complete wetting of the hard wall by the crystal in [111] orientation might occur ($\gamma_{wf}(\eta_t)-\gamma_{wc}^{111}(\eta_t)>\gamma^{111}$), but a finite, small contact angle cannot be excluded from the errorbars. We mention in passing that the difficulties in extracting interface tension with reliable errorbars might be substantial: as an example we mention the values for the interfacial {\em stiffness} $\tilde{\gamma}^{100}$ obtained in different simulations using the capillary wave fluctuation method. Laird and Davidchack \cite{27a} obtain $\tilde{\gamma}^{100}=0.44\pm0.03$ (using thin slabs) whereas Zykova-Timan et al. obtain  $\tilde{\gamma}^{100}=0.49 \pm 0.02$ (using thick slabs). These stiffnesses include the anisotropy of the interfacial tension in an amplified manner but apparently depend on the simulation geometry.

From Fig.~\ref{fig7} we conclude that changing the wall potential from the hard wall case \{Eq.~\ref{eq1}\} to the WCA case \{Eq.~\ref{eq2}\} has little effect on the wetting properties of the wall, since the difference $\gamma_{wf}(\eta_t)-\gamma_{wc}^{111}(\eta_t)$ is independent of $\epsilon$, at least within the statistical errors of our calculation, and moreover is almost identical to the HW results. Thus, although the variation of $\epsilon$ from $\epsilon=0.25$ to $\epsilon = 4.0$ enhances $\gamma_{wf}(\eta_t)$ by about 0.5, the increase of $\gamma_{wf}(\eta_t)$ is almost identical to the increase of $\gamma_{wc}(\eta_t)$, and hence one cannot reach a wetting transition (and then vary the contact angle) by varying $\epsilon$.

\section{Conclusions}
In this work, the effects of confining walls on a hard sphere fluid were studied over a wide range of packing fractions, including also the regime of the transition to the solid crystalline phase. The effect of the wall was described by using a WCA potential \{Eq.~\ref{eq2}\} acting on the fluid particles, but for comparison also a hard wall potential \{Eq.~\ref{eq1}\} was chosen. The main interest of this paper was a comparative study of various methods to obtain the surface excess free energy and the surface excess density, applying both Monte Carlo (MC) methods and DFT calculations. We found very good agreement between all approaches in the fluid phase for not too large packing fractions ($\eta <0.4$), irrespective of the choice of the wall-fluid potential that was used. For $\eta >0.35$, systematic discrepancies between the MC and DFT results for the surface excess density were found, which presumably should be attributed to the fact that for high densities in the fluid nontrivial correlations between the fluid particles beyond the nearest neighbor shell develop, which are no longer described by DFT with very high accuracy. However, DFT describes very accurately the density distribution very close to the walls, and since this controls the wall-fluid surface tension, the latter is very accurately predicted by DFT (Fig. ~\ref{fig5}b).

The application of MC methods for $\eta >0.4$ also becomes increasingly difficult - the pronounced layering that occurs makes the procedures that we used sensitive to finite size effects both with respect to $D$ (the regions disturbed by both walls start to interact) and with respect to $L$ (when a precursor of a crystalline wetting layer occurs at a wall, the crystalline planes exhibit an in-plane triangular lattice structure, which exhibits a mismatch with an $L \times L$ cross-section due to the periodic boundary conditions). This problem occurs a fortiori in the solid phase (where also $D$ needs to be chosen such that elastic distortion of the crystal in z-direction is avoided). Thus, our study is clearly  a feasibility study only, and more work will be required to ascertain the true behavior occurring in the thermodynamic limit. We recall that for the case of hard walls, that we have included for comparison, many studies by different methods were indispensable to reach the current level of understanding.

One motivation of the present work was also to possibly control the difference $\gamma_{wf}(\epsilon) - \gamma_{wc}(\epsilon)$ at the bulk fluid-solid transition by varying $\epsilon$, in order to allow a convenient study of a wetting transition at crystallization. However, unfortunately the variation of this difference with $\epsilon$ is rather weak, and the system stays in the region of complete wetting (zero contact angle) or in the regime of small nonzero contact angles, so one cannot reach states deep in the incomplete wetting regime in this way. Nevertheless, our calculations could be useful to understand experiments were one uses walls coated with polymer brushes containing hard-sphere like colloidal dispersions.

\underline{Acknowledgements}: We acknowledge support by the Deutsche Forschungsgemeinschaft (DFG) under grants No Bi 314/19-2, SCHI 853/2-2, and SFB TR6 and the JSC for a grant of computer time.

\appendix 
\section{Determination of surface tensions using density functional theory}
The equilibrium solvent density profile $\rho(\vect r) \equiv \req(\vect r)$
can be  determined directly from the basic equations
of density functional theory. The grand potential functional is given by
\bea
 \label{eq:omegadef}
  \Omega[\rho] &=& {\cal F}^{\rm id}[\rho] +  {\cal F}^{\rm ex}[\rho] - \int
   d\vect r(\mu - V^{\rm ext}(\vect r)) \;,
\eea
where ${\cal F}^{\rm id}$ and ${\cal F}^{\rm ex}$ denote the ideal and excess free
energy functionals of the solvent. The chemical potential in the hard sphere fluid is denoted by $\mu$ and
and the wall (hard or soft)  defines the external potential $V^{\rm ext}$ (given by
Eqs.~\ref{eq1}, \ref{eq2}) and which depends only on the Cartesian
coordinate $z$.
The exact form of the ideal part of the free energy is given by
\bea
  \beta {\cal F}^{\rm id}[\rho] &=& \int d\vect r\beta f^{\rm id}(\vect r) = \int d^3r\, \rho(\vect r )\left(
 \ln[ \rho(\vect r) \Lambda^3] -1 \right)\;.
\eea
Here, $\Lambda$ is the de--Broglie wavelength and $\beta=1/(\kt)$ is the inverse temperature.
The equilibrium density profile $\req(\vect r)$ for the solvent at
chemical potential $\mu=\beta^{-1}\ln(\rho_b\, \Lambda^3) + \mex$
(corresponding to the bulk density $\rho_b$)
is found by minimizing the grand potential in Eq.~(\ref{eq:omegadef}):
\bea
 \label{eq:rhoeq_main}
  \ln \frac{\req(z)}{\rho_s} + \beta V^{\rm ext}(z) =
   -\beta\frac{\delta {\cal F}^{\rm ex}[\req]} {\delta \rho(z)}+ \beta\mex  \;.
\eea

For an explicit solution, it is necessary to specify the excess part of the
free energy.
 We
employ fundamental measure functionals which represent the most precise functionals
for the hard sphere fluid. Specifically we employ:
\bea
 \label{eq:fhs}
 {\cal F}^{\rm ex} &=& \int d\vect r \, f^{\rm ex}( \{\vect n[\rho (\vect r)]\} ) \;,  \\
  \beta  f^{\rm ex}( \{\vect n[\rho (\vect r)]\} ) &=&   -n_0\,\ln(1-n_3) +
      \varphi_1(n_3)\frac{n_1 n_2-\vect n_1 \cdot \vect n_2}{1-n_3} + \nonumber \\
  & &      \varphi_2(n_3)\;\frac{n_2^3-3n_2\, \vect n_2\cdot \vect n_2 +
    \alpha_{\rm T}\;
   \frac{\di 9}{\di 2}\left( \vect n_2 \cdot n_{\rm T} \cdot \vect n_2 - {\rm Tr}  n_{\rm T}^3   \right) }{24\pi(1-n_3)^2}\;.
   \nonumber \\
 \varphi_1 & = & 1 + \frac{2n_3-n_3^2 + 2(1-n_3) \ln(1-n_3 )}{3n_3}  \nonumber \\
 \varphi_2 & = & 1 - \frac{2n_3-3n_3^2 + 2n_3^3 + 2(1-n_3 )^2 \ln(1-n_3 ) }
                          {3 n_3^2}  \nonumber
\eea
Here, $f^{\rm ex}$ is  a
free energy density which is a function of a set of weighted densities
$\{\vect n (\vect r)\} = \{ n_0,n_1,n_2,n_3,\vect n_1,\vect n_2, n_{\rm T}\}$
with four scalar,  two vector  and one tensorial weighted densities. These are related to the
density profile $\rho(\vect r)$ by
$n_\alpha{(\vect r)} =  \int d\vect r' \rho(\vect r') \, w^\alpha(\vect r- \vect r')$.
The weight functions,
$\{\vect w(\vect r)\} = \{w^0,w^1,w^2,w^3, \vect w^1,\vect w^2, w_{\rm T}\}$,
depend on the hard sphere radius $R=\sigma/2$
as follows:
\bea
 w^3 = \theta(R-|\vect r|)\;, \qquad
 w^2 = \delta(R-|\vect r|)\;, \qquad w^1 = \frac{w^2} {4\pi R}\;,
  \qquad w^0 = \frac{w^2}{4\pi R^2}\;,
  \nonumber\\
 \vect w^2 =\frac{\vect r}{|\vect r|}\delta(R-|\vect r|)\;,
 \qquad \vect w^1 = \frac{\vect w^2}{4\pi R} \;,
 \qquad (w_{\rm T})_{ij} = \left( \frac{r_i r_j}{r^2} - \frac{\delta_{ij}}{3}\right) \delta(R-|\vect r|)\;.
\eea
Setting $\alpha_{\rm T}=0$ in Eq.~(\ref{eq:fhs}) corresponds to neglecting the tensorial weighted density.
This is the White Bear II (WBII) functional derived in Ref.~\cite{Han06}. This functional is most consistent with
restrictions imposed by morphological thermodynamics \cite{Koe04}, see below for a discussion what this means
for the hard wall  surface  tension. Setting $\alpha_{\rm T}=1$ corresponds to the tensor modification
(originally introduced in Ref.~\cite{Tar00})
of WBII (WBII--T) which facilitates the hard sphere crystal description. Coexistence densities, bulk crystal free energies,
density anisotropies in the unit cell and vacancy concentrations are described very well using WBII--T \cite{Oet10}.

From the equilibrium density profiles $\req(z)$, the surface tension can be determined as the excess
over bulk grand potential:
\bea
 \label{eq:gamma_req}
  \gamma[\req] = \int_{z_0}^\infty dz \left[f^{\rm id}[\req(z)] + f^{\rm ex}[\req(z)] -(\mu - V^{\rm ext}(z))\req(z)
       - \omega_b   \right] \;,
\eea
where $z_0$ denotes the location of the wall and the grand potential density in the bulk is given by the
negative pressure, $\omega_b=-p$. Both the WBII and the WBII--T functional are consistent with the Carnahan--Starling
equation for $p$.

In the case of a hard wall, the surface tension can be determined from a scaled particle argument
\cite{Han06,Bot09} as follows:
\bea
 \label{eq:gamma_spt}
 \gamma_{\rm SP} = \left.\frac{\partial f^{\rm ex}}{\partial  n_2 }\right|_{\{\vect n\}=\{\vect n_b\}} =
     -\frac{\ln(1-\eta_b)}{\pi} + \frac{\eta_b(2+3\eta_b-2\eta_b^2)}{\pi(1-\eta_b)^2} \;.
\eea
This surface tension is taken with respect to the wall position $z_0$ being at the physical wall and not at the
surface of exclusion $z_0'=z_0+\sigma/2$ where the wall potential jumps from infinity to zero.
Here, the derivative of $f^{\rm ex}$ has to be evaluated with the bulk values for the set of weighted
densities: $n_{3,b}=\eta_b$, $n_{2,b}=6/\sigma\,\eta_b$, $n_{1,b}=3/(\pi\sigma^2)\,\eta_b$, $n_{0,b}=6/(\pi\sigma^3)\,\eta_b$,
$\vect n_{1,b} = \vect n_{2,b} = n_{{\rm T},b}=0$ with $\eta_b=\sigma^3\pi/6\,\rho_b$ denoting the bulk packing fraction.
For a consistent functional, both expressions for the surface tension \ref{eq:gamma_req} and \ref{eq:gamma_spt} should
agree. The WBII functional is very consistent in this respect, as illustrated in Tab.~\ref{tab:gamma}, and the WBII--T functional
is only slightly less consistent. For packing fractions larger than 0.45 (close to freezing) the inconsistency becomes noticeable,
this is also where
we observe the largest deviations from the simulation results. The analytical $\gamma_{\rm SP}$ is still closest to the simulation results.

For soft walls, no analytical result can be derived. One would extrapolate from the hard wall results that $\gamma[\req]$ from the
WBII functional will give slightly better results than $\gamma[\req]$ from the WBII--T functional. This is indeed what we have observed
in comparison to the simulations.

\clearpage

\begin{table}
 \begin{tabular}{llll} \hline \hline
   $\eta_b\qquad\qquad$ & $\beta\sigma^2\,\gamma_{\rm SP}\qquad$ & $\beta\sigma^2\,\gamma[\req]\qquad$  & $\beta\sigma^2\,\gamma[\req]\qquad$  \\
     & & (WBII) & (WBII--T) \\ \hline
    0.1 & 0.1231 & 0.1232 & 0.1232 \\
    0.2 & 0.3217 & 0.3218 & 0.3219 \\
    0.3 & 0.6436 & 0.6419 & 0.6436 \\
    0.4 & 1.181 & 1.177 & 1.187 \\
    0.45 & 1.585 & 1.589 & 1.610 \\
    0.47 & 1.783 & 1.798 & 1.825 \\
    0.49 & 2.007 & 2.040 & 2.074 \\ \hline \hline
  \end{tabular}
 \caption{\label{tab:gamma} Comparison of surface tension $\gamma_{\rm SP}$ vs. $\gamma[\req]$ of hard spheres against a hard wall
  for various bulk packing fractions up to freezing. }
\end{table}

\end{document}